\documentclass[runningheads]{llncs}
\usepackage{graphicx}
\usepackage{array,multirow}
\usepackage{booktabs}       
\usepackage{dirtytalk}
\usepackage{amsfonts}       
\usepackage{nicefrac}       
\usepackage{microtype}      
\usepackage{amsmath}
\usepackage{graphicx}
\usepackage{amssymb}
\usepackage[utf8]{inputenc} 
\usepackage[T1]{fontenc}    
\usepackage{hyperref}       
\usepackage{url}            
\usepackage{booktabs}       
\usepackage{amsfonts}       
\usepackage{lipsum}
\usepackage{graphicx}
\usepackage{authblk}
\usepackage{amssymb}
\usepackage{mathtools}
\usepackage{color}
\usepackage{lmodern}

\makeatletter
\newcommand{\printfnsymbol}[1]{%
  \textsuperscript{\@fnsymbol{#1}}%
}
\makeatother

\begin{document}
\title{Body Fat Estimation from Surface Meshes using Graph Neural Networks}
%
%
\author{Tamara T. Mueller\thanks{These authors contributed equally to this work.}\inst{1,2} \and
Siyu Zhou\printfnsymbol{1}\inst{1} \and
Sophie Starck\inst{1} \and
Friederike Jungmann\inst{2} \and
Alexander Ziller \inst{1,2} \and
Orhun Aksoy\inst{1} \and
Danylo Movchan\inst{1} \and
Rickmer Braren\inst{2} \and
Georgios Kaissis\inst{1,2,4} \and
Daniel Rueckert\inst{1,3}
}

\authorrunning{Mueller and Zhou et al.}
%
\institute{Institute for AI in Medicine and Healthcare, Faculty of Informatics, Technical University of Munich \and
Department of Diagnostic and Interventional Radiology, Faculty of Medicine, Technical University of Munich \and
Department of Computing, Imperial College London \and
Institute for Machine Learning in Biomedical Imaging, Helmholtz-Zentrum Munich \\
\email{tamara.mueller@tum.de}
}

\maketitle
\begin{abstract}
    Body fat volume and distribution can be a strong indication for a person's overall health and the risk for developing diseases like type 2 diabetes and cardiovascular diseases. 
    Frequently used measures for fat estimation are the body mass index (BMI), waist circumference, or the waist-hip-ratio. 
    However, those are rather imprecise measures that do not allow for a discrimination between different types of fat or between fat and muscle tissue.
    The estimation of visceral (VAT) and abdominal subcutaneous (ASAT) adipose tissue volume has shown to be a more accurate measure for named risk factors. 
    In this work, we show that triangulated body surface meshes can be used to accurately predict VAT and ASAT volumes using graph neural networks.
    Our methods achieve high performance while reducing training time and required resources compared to state-of-the-art convolutional neural networks in this area. 
    We furthermore envision this method to be applicable to cheaper and easily accessible medical surface scans instead of expensive medical images.
\end{abstract}

\section{Introduction}

\begin{figure}[t]
    \centering
    \includegraphics[width=1\textwidth]{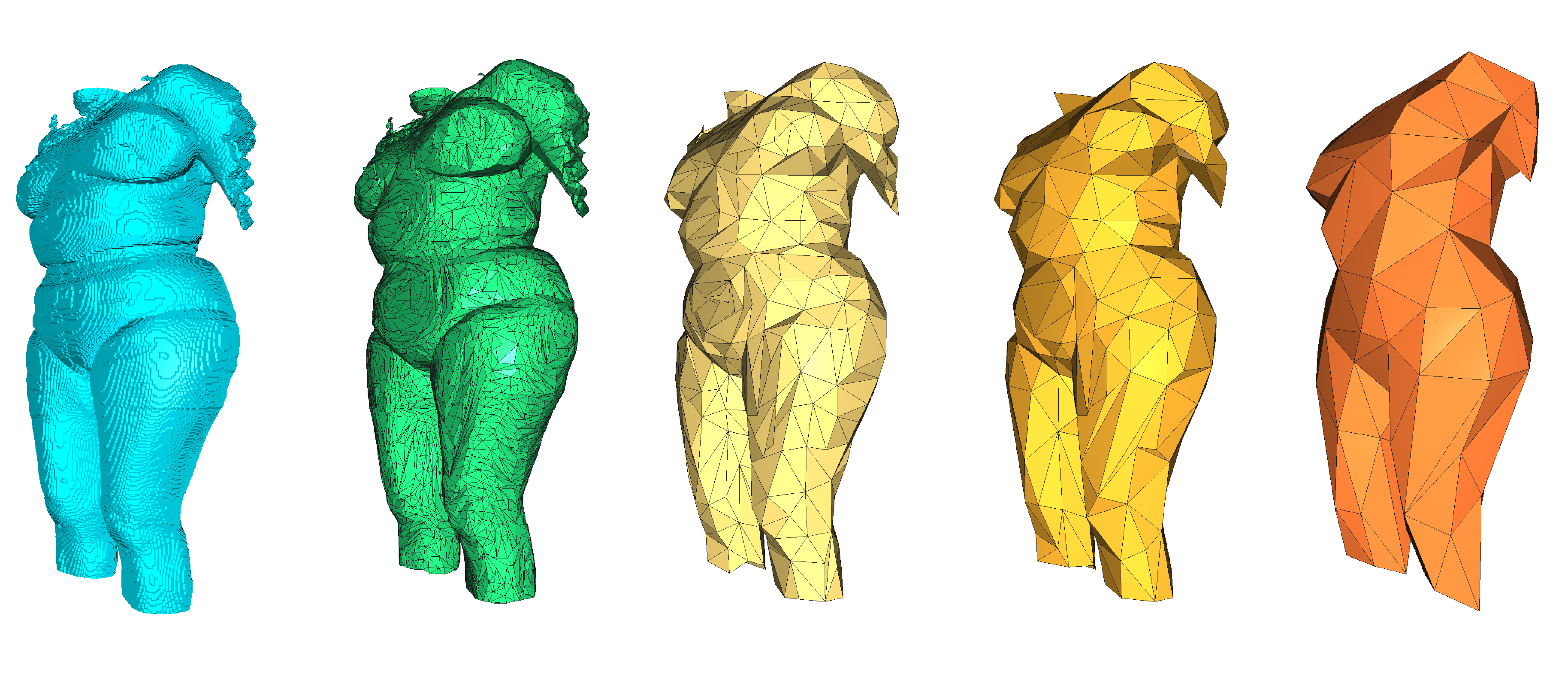}
    \caption{Visualisation of body surface meshes at different decimation rates; The most left mesh shows the original mesh, then left to right are visualisations of decimated meshes with ten thousand, one thousand, five hundred and two hundred faces.}
    \label{fig:surfaces}
\end{figure}

The estimation of body composition measures refers to the qualification and quantification of different tissue types in the body as well as the estimation of their distribution throughout the body. These measures can function as risk factors of individuals and be an indicator for health and mortality risk  \cite{afshin2017health,calle2003overweight}. One component of body composition analysis is the estimation of fatty tissue volume in the body. 
The strong correlation between body composition and disease risk has lead to a routine examination of measures indicating body composition in medical exams. The body mass index (BMI), for example, measures the ratio between a person's weight and height and has been shown to be an indicator for developing cardiovascular diseases, type 2 diabetes, as well as overall mortality \cite{kivimaki2017overweight,calle2003overweight,anderson2020body,larsson2020body}. Additionally, the waist circumference and waist-hip-ratio can be used as an indication for body fat distribution \cite{neeland2019visceral,song2013comparison,jacobs2010waist,baioumi2019comparing}. These metrics are easy, fast, and cheap to assess. However, they have strong limitations. They are imprecise as they do not allow for a more accurate assessment of the distribution of body fat or to differentiate between weight that stems from muscle or fat tissue. 
Understanding the specific differences between different types of fatty tissue and their impact on health risks is crucial for accurately assessing an individual's risk factors and enabling personalised medical care.
Towards this goal, several works have investigated methods to identify variations of fat distribution in the body and the quantification of fatty tissues \cite{xie2015accurate,klarqvist2022silhouette}. 

Body fat can be divided into different types of fat. Two commonly investigated types are \textit{visceral fat} (VAT), which surrounds the abdominal organs, and \textit{abdominal subcutaneous fat} (ASAT), which is located beneath the skin. Studies have shown that especially visceral fat can have a negative impact on a person's health \cite{matsuzawa1995visceral,bergman2006visceral,shuster2012clinical}. Therefore, a separate analysis of VAT and ASAT is an important step towards gaining accurate insights into  body composition. Several works have investigated a precise estimation of VAT and ASAT volumes from medical images, like magnetic resonance (MR) \cite{klarqvist2022silhouette} and computed tomography (CT) images \cite{hemke2020deep}, dual-energy X-ray absorptiometry (DXA) assessment \cite{messina2020body}, or ultrasound imaging \cite{bazzocchi2016ultrasound}. Deep learning techniques have shown promising results in analysing these medical images in order to estimate body composition values \cite{klarqvist2022silhouette,hemke2020deep,wang2020artificial,nowak2020fully}.

In this work, 
we perform VAT and ASAT volume prediction from full body triangulated surface meshes using graph neural networks (GNNs). We show that GNNs allow to utilise the full 3D data at hand, thereby achieving better results than state-of-the-art convolutional neural networks (CNNs) on 2D silhouettes, while requiring significantly less training time and therefore resources. Both ours and related work, such as \cite{klarqvist2022silhouette}, use data extracted from MR images. 
However, MR imaging is a very expensive technique, which is highly unequally distributed around the globe. The access to MR scanners in lower income countries is much more limited \cite{geethanath2019accessible}. Furthermore, the acquisition of MR images is time consuming and very unlikely to be used for routine exams. Given the light computational weight and fast nature of our method, we envision it to  be applied to data acquired from much simpler surface scans in the future and enable an incorporation into routine medical examination.

\section{Background and Related Work}
In the following, we summarise related works on body fat estimation from medical (and non-medical) images, define triangulated meshes and the concept of graph neural networks and show some of their application to medical data, with a focus on surface meshes.

\subsection{Body Fat Estimation from Medical Imaging}

Body fat estimation has been part of routine medical assessments for decades through the analysis of simple measurements such as BMI or waist circumference \cite{fan2022body}. 
However, more elaborate ways such as using proxy variables derived from medical images, like dual energy X-ray absorptiometry (DXA), CT or MR images, have achieved more accurate results. Multiple studies have successfully assessed patient body composition based upon DXA \cite{harty2020novel,direk2013relationship,messina2020body}. Hemke et al. \cite{hemke2020deep} and Nowak et al. \cite{nowak2020fully} show successful utilisation of CT images for body composition assessment. Works like \cite{Kstner2020FullyAA} use segmentation algorithms to identify fatty tissue in MR scans, from which body composition values can be derived.  
Tian et al. \cite{tian2020predicting} estimate body composition measures based on 2D photography, not even requiring medical imaging techniques. 
Many of these approaches focus on predicting specific types of adipose tissue \cite{LINDER2020109184,LU2023511,klarqvist2022silhouette,Kstner2020FullyAA}. 
One idea, that has been followed by several works is the utilisation of silhouettes, a binary 2D projection of the outline of the body extracted from images. Xie et al. \cite{xie2015accurate} use silhouettes generated from DXA whole-body scans to estimate shape variations and Klarqvist et al. \cite{klarqvist2022silhouette} use silhouettes derived from MR Images for VAT and ASAT volume estimation using CNNs. The latter use two-dimensional coronal and sagittal silhouettes of the body outline and predict VAT and ASAT volume using convolutional neural networks. The silhouettes are extracted from the full-body magnetic resonance (MR) scans of the UK Biobank dataset \cite{ukbb}. In our work, we propose to switch from full medical images or binary silhouettes to surface meshes for fat volume prediction, which allows to integrate the full potential of the 3D surface into deep learning methods, while using the light-weight and fast method of graph neural networks (GNNs).


\subsection{Triangulated Meshes}
In this work, we use triangulated surface meshes of the body outline. A mesh structure can be interpreted as a specific 3D representation of a graph. A graph $\mathit{G}:= (\mathit{V}, \mathit{E})$ is defined by a set of nodes $\mathit{V}$ and a set of edges $\mathit{E}$, connecting pairs of nodes. The nodes usually contain node features, which can be summarised in a node features matrix $\mathbf{X}$. A triangulated mesh $\mathit{M}$ has the same structure, commonly holding the 3D coordinates of the nodes as node features. All edges form triangular faces that define the surface of the object of interest --in our case: body surfaces. A visualisation of such meshes can be found in Figure \ref{fig:surfaces}. 

\subsection{Graph Neural Networks}
 Graph neural networks have opened the field of deep learning to non-Euclidean data structures such as graphs and meshes \cite{bronstein2017geometric}. Since their introduction by \cite{gori2005new} and \cite{scarselli2008graph}, they have been utilised in various domains, including medical research \cite{ahmedt2021graph,ding2022graph}. Graphs are, for example, frequently used for representations of brain graphs \cite{bessadok2022graph}, research in drug discovery \cite{bonner2022review}, or bioinformatics \cite{yi2022graph,zhang2021graph}. One native data structure that benefits from the utilisation of graph neural networks are surface meshes \cite{bronstein2017geometric}.
 GNNs on mesh datasets have also advanced research in the medical domain such as brain morphology estimation \cite{azcona2020interpretation}, which can be used for Alzheimer's disease classification, or for the predicting of soft tissue deformation in image–guided neurosurgery \cite{salehi2022physgnn}. 
 

In general, GNNs follow a so-called message passing scheme, where node features are aggregated among neighbourhoods, following the underlying graph structure \cite{kipf2016semi,chiang2019cluster,huang2020combining,kong2020flag}. This way, after each iteration, a new embedding for the node features is learned. In this work, we use Graph SAGE \cite{hamilton2017inductive} convolutions, which were designed for applications on large graphs. The mean aggregator architecture for a node $v \in \mathcal{V}$ at step $k$ is defined as follows: 

\begin{equation}
    h_v^{k} = \sigma \left (  \mathbf{W} \cdot \text{MEAN} (\{ h_v^{k-1} \} \cup \{  h_u^{k-1}, \forall u \in \mathcal{N}_v \} ) \right ).
\end{equation}

$\mathcal{N}_v$ is the neighbourhood of node $v$, $\mathbf{W}$ is a learnable weight matrix, and $\text{MEAN}$ the mean aggregator, which combines the node features of $v$ at the previous step and the node features of $v$'s neighbours.

\section{Methods}
We construct three different model architectures: (a) a graph neural network, (b) a simple convolutional neural network (CNN), and (c) a DenseNet and compare their performance. All models are trained using the Adam optimiser \cite{Kingma2014AdamAM} and Shrinkage loss \cite{lu2018deep} and all results reported are cross-validated based on a $5$-fold data split. We use a Quadro RTX $8\,000$ GPU for our experiments and all models predict both targets --VAT and ASAT-- with the same network, following the approach from \cite{klarqvist2022silhouette}.

\subsubsection{GNN Architecutre}
We perform a whole-graph regression task on the input meshes. The model architecture consists of a three-layer GNN with SAGE graph convolutions \cite{hamilton2017inductive} and batch normalisation layers, followed by a max aggregation and a three-layer multi-layer perceptron (MLP).
Hyperparameters such as learning rate and GNN layers are selected by manual tuning. All GNNs are trained for $150$ epochs.

\subsubsection{CNN Architecture}
In order to compare our results to the work by Klarqvist et al. \cite{klarqvist2022silhouette}, we also train a DenseNet and a simpler CNN on the silhouette data. DenseNet is a CNN which is more densely connected, where each layer takes all previous outputs as an input. For our DenseNet implementation, we follow the architecture in \cite{klarqvist2022silhouette}. We additionally construct a simpler CNN architecture that consists of three 2D convolutions, followed by a three-layer MLP, matching the design of the graph neural networks. Both convolutional networks are trained for $20$ epochs on a 2D input image, that consist of a sagittal and a coronal view of the binary silhouette masks of the MR images, following the pipeline in \cite{klarqvist2022silhouette}.

\section{Experiments and Results}
\begin{figure}[t]
    \centering
    \includegraphics[width=\textwidth]{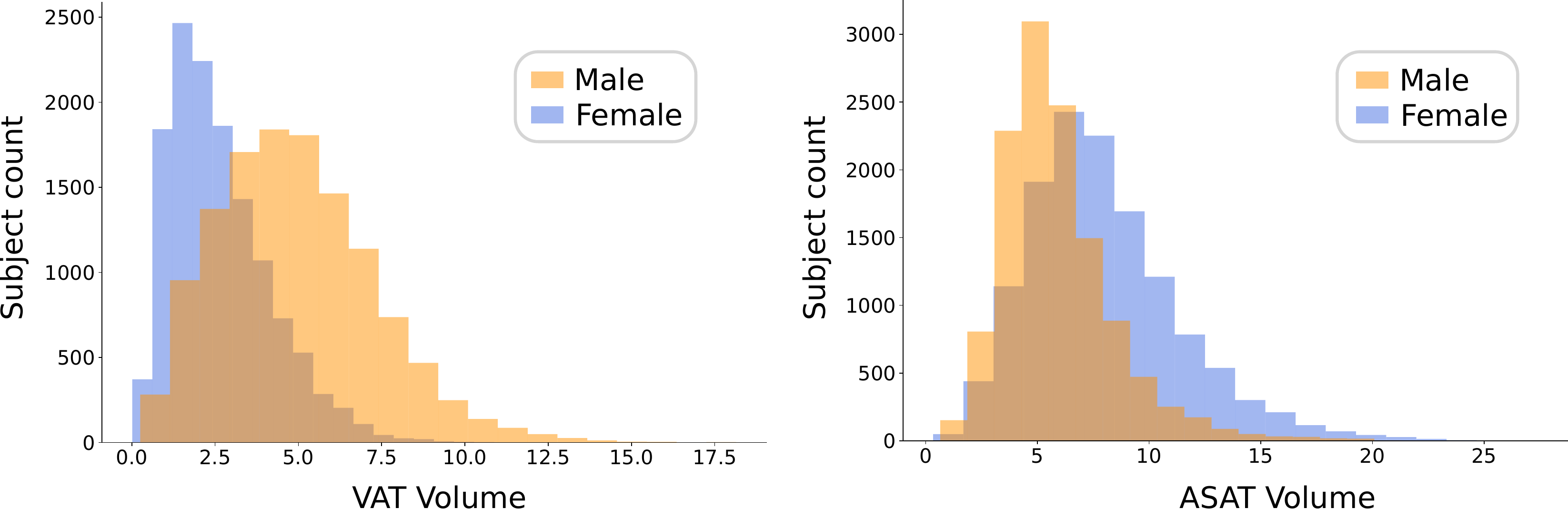}
    \caption{Distribution of VAT (left) and ASAT (right) volume of male and female subjects in the cohort. Male subjects tend to have more VAT volume, whereas female subjects tend to have more ASAT volume.}
    \label{fig:vat_asat_distributions}
\end{figure}

We use a subset of the UK Biobank dataset \cite{ukbb}, which is a large-scale medical database. It contains a variety of imaging data, genetics, and life-style information from almost $65\,000$ subjects and was acquired in the United Kingdom. In this work, we use the neck-to-knee magnetic resonance images of a subset of $25\,298$ subjects, for which the labels are available ($12\,210$ male and $13\,088$ female). The mean age of this cohort is $62.95$ years. The VAT and ASAT distributions of male and female subjects are visualised in Figure \ref{fig:vat_asat_distributions}. We can see that female subjects tend to have a higher ASAT volume, whereas male subjects tend to have more VAT. As labels, we used the reported VAT and ASAT volumes in the UK Biobank (field IDs: $22407$ and $22408$).

\subsection{Data Processing}
The experiments in this work are performed on triangulated body surface meshes that are extracted from the neck-to-knee MR images from the UK Biobank \cite{petersen2013imaging}. These were acquired in stations and merged through stitching \cite{lavdas2019machine}. In order to extract the surface meshes, we first perform an algorithmic whole-body segmentation by a succession of morphological operations on the stitched MR scans. We then convert these segmentations into surface meshes using the marching cubes algorithm \cite{lorensen1998marching} and the open3d library \cite{Zhou2018}. 
In order to investigate how much the surface meshes can be simplified, we decimate them into meshes consisting of different numbers of faces. We use meshes with $10\,000$, $5\,000$, $1\,000$, $500$, $200$, and $100$ faces. The number of nodes is always half the number of faces, following Euler's formula for triangular meshes \cite{euler1740summis}. Subsequently, the meshes are registered into a common coordinate system, using the iterative closest point algorithm \cite{4767965}. As a reference subject, the most average subject in the dataset was selected based on height, weight, and age.
The resulting decimated and registered surface meshes are then used for graph learning. Figure \ref{fig:surfaces} shows an example of a body surface mesh at different decimation rates.


\subsection{Results}

\begin{figure}[t]
    \centering
    \includegraphics[width=\textwidth]{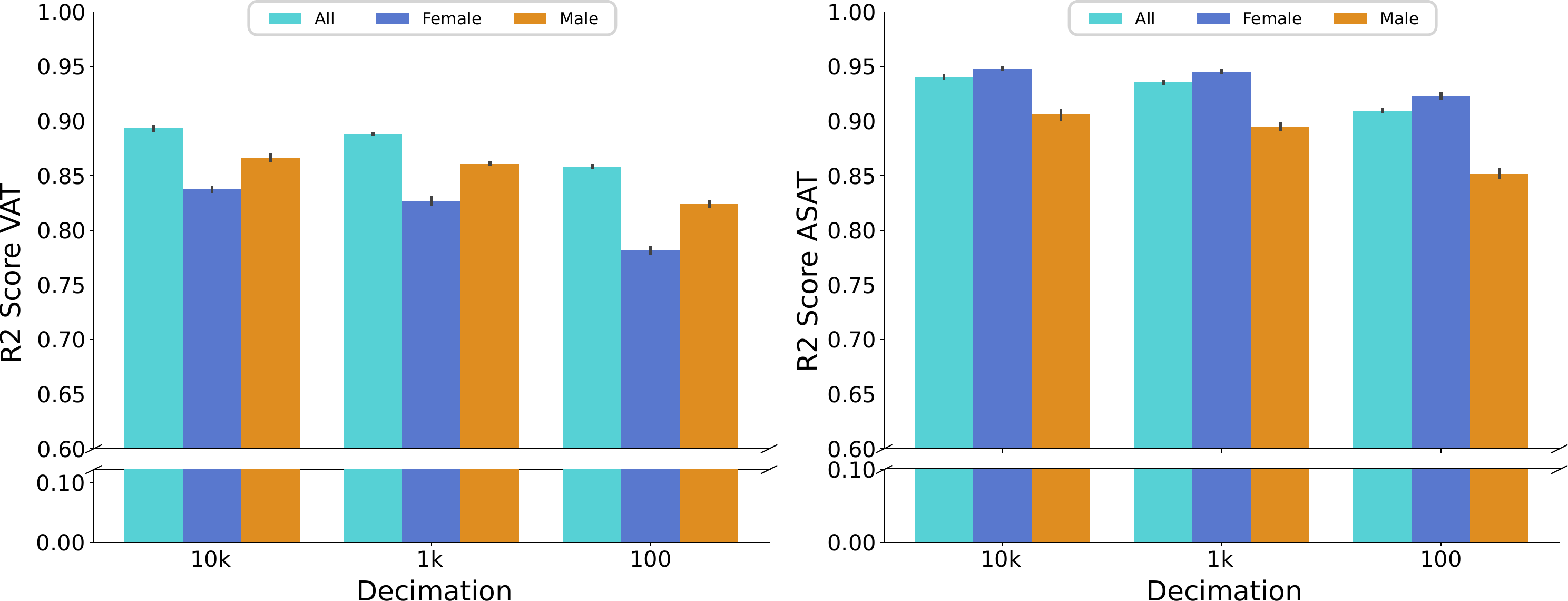}
    \caption{R2 score results of VAT (left) and ASAT (right) predictions for all subjects, only males, and only females.}
    \label{fig:results}
\end{figure}

\begin{table*}[!ht]
\centering
\caption{Results for \textbf{VAT} and \textbf{ASAT} volume estimation; We report the R2 scores on the test set with standard deviations based on 5-fold cross validation, as well as the training times of the full training in minutes.}
\addtolength{\tabcolsep}{5pt}
\begin{tabular}{lllcc}
\toprule[0.5pt]
\textbf{Tissue} & \textbf{Model} & \textbf{Decim.} & \textbf{Test R2} & \textbf{Time (min)} \\
\toprule[0.8pt]
VAT & GNN (ours) & 100  & 0.858 $\pm$ 0.001  & \textbf{8.36} \\
& & 200 & 0.872 $\pm$ 0.001  &  8.63  \\
& & 500 & 0.882 $\pm$ 0.001  &  9.01  \\
&  & 1k & 0.888 $\pm$ 0.001  &  10.11 \\
& & 5k & \textbf{0.893 $\pm$ 0.002}  & 22.36 \\
& & 10k & 0.893 $\pm$ 0.003  & 37.75 \\
\cmidrule{2-5}
& CNN (ours) & - & 0.874 $\pm$ 0.001 & 16.20 \\
\cmidrule{2-5}
& DenseNet & - & 0.878 $\pm$ 0.004 &  95.79 \\
\midrule
ASAT & GNN (ours) & 100 & 0.909 $\pm$ 0.001 & \textbf{8.36} \\
 &  & 200 & 0.921 $\pm$ 0.002 &  8.63\\
 &  & 500 & 0.931 $\pm$ 0.001 &  9.01\\
 &  & 1k & 0.935 $\pm$ 0.002 & 10.11 \\
 & & 5k & 0.938 $\pm$ 0.000 &  22.36 \\
 & & 10k & \textbf{0.941 $\pm$ 0.002} &  37.75  \\
\cmidrule{2-5}
 & CNN (ours) & - & 0.921 $\pm$ 0.002 & 16.20\\
 \cmidrule{2-5}
& DenseNet & - & 0.934 $\pm$ 0.002 & 95.79 \\
\bottomrule
\end{tabular}
\label{tab:asat_vat_results}
\end{table*}

Table \ref{tab:asat_vat_results} summarises the results of the GNNs and CNNs for ASAT and VAT volume prediction. We report the 5-fold cross-validation results on the test set of the best performing models, evaluated on the validation loss.
We compare the results of our graph neural networks (GNNs) with the results achieved by the DenseNet from \cite{klarqvist2022silhouette} and the results of a simpler CNN (which we call \textit{CNN} in the tables).
We furthermore report the training times of all models, measured by the full training process for $150$ and $20$ epochs for GNNs and CNNs, respectively.
All GNNs are trained on the body surface meshes, whereas the CNNs are trained on the silhouettes, following the approach proposed in \cite{klarqvist2022silhouette}.
We evaluate the GNNs on body surface meshes at different decimation rates of ten thousand, five thousand, one thousand, $500$, $200$, and $100$ faces per mesh (see Figure \ref{fig:surfaces} for a visualisation of some of these decimated meshes). The best test performances are highlighted in bold, so are the shortest training times. 
We can see that the simpler CNN architecture almost matches performance of the DenseNet proposed by \cite{klarqvist2022silhouette}, while requiring less training time. The GNNs outperform the CNN and the DenseNet, when the utilised meshes are not heavily decimated. But even highly decimated surface meshes with one hundred faces, only result in minor performance loss while requiring less than ten times less training time compared to the DenseNet. We envision the utilisation of the surface meshes and graph neural networks to allow for more efficient model training and the utilisation of the full 3D structure of the body, while keeping resource requirements low.

\begin{table*}[ht!]
\centering
\addtolength{\tabcolsep}{5pt}
\caption{Results of \textbf{VAT} and \textbf{ASAT} volume prediction split by subject sex; all reported values are R2 scores on the test set, cross-validated across 5 folds.}
\begin{tabular}{lllcc}
    \toprule
    \textbf{Fat tissue} &
    \textbf{Model} &
    \textbf{Decimation} &
    \textbf{Female R2} &
    \textbf{Male R2} \\
    \midrule
    VAT & GNN (ours) & 100 &  0.782 $\pm$ 0.004 & 0.824 $\pm$ 0.003 \\
    &  & 200 &  0.804 $\pm$ 0.006 & 0.840 $\pm$ 0.003 \\
    &  & 500 &  0.815 $\pm$ 0.008& 0.854 $\pm$ 0.003 \\
    &  & 1k &  0.827 $\pm$ 0.004& 0.861 $\pm$ 0.001 \\
    &  & 5k &  0.831 $\pm$ 0.006& \textbf{0.868 $\pm$ 0.002 }\\
    &  & 10k &  \textbf{0.837 $\pm$ 0.002} & 0.867 $\pm$ 0.004  \\
    \cmidrule{2-5}
    & CNN (ours) & - & 0.804 $\pm$ 0.003 & 0.845 $\pm$ 0.002 \\
    \cmidrule{2-5}
    & DenseNet & - & 0.811 $\pm$ 0.006 & 0.849 $\pm$ 0.006\\
   \midrule
   ASAT & GNN (ours) & 100 & 0.923 $\pm$ 0.003  & 0.852 $\pm$ 0.004  \\
    &  & 200 & 0.934 $\pm$ 0.001 & 0.870 $\pm$ 0.006 \\
    &  & 500 & 0.940 $\pm$ 0.002 & 0.890 $\pm$ 0.002 \\
    &  & 1k & 0.945 $\pm$ 0.001 & 0.895 $\pm$ 0.004 \\
    &  & 5k & 0.945 $\pm$ 0.000 & 0.903 $\pm$ 0.002 \\
    &  & 10k &  \textbf{0.948 $\pm$ 0.001} & \textbf{0.906 $\pm$ 0.005} \\
    \cmidrule{2-5}
    & CNN (ours) & - & 0.934 $\pm$ 0.002 & 0.870 $\pm$ 0.002 \\
    \cmidrule{2-5}
    & DenseNet & - & 0.944 $\pm$ 0.001 & 0.891 $\pm$ 0.003 \\
    \bottomrule
\end{tabular}
\label{tab:sexbased}
\end{table*}

Male and female subjects show different distributions in VAT and ASAT volume. While male subjects tend to have more VAT, females tend to have more ASAT. Figure \ref{fig:vat_asat_distributions} shows the distributions of the fat volumes of the two sex groups. 
We therefore compare the results of our method for female and male subjects separately. Table \ref{tab:sexbased} summarises the results of all GNNs and CNNs for VAT and ASAT volume prediction split by sex. The best performing model for each fat type and sex is highlighted in bold. We can see that the predictions of VAT volume tends to be better on male subjects whereas the prediction of ASAT volume achieves slightly higher scores for the female subject. The GNNs, however, seem to show a slightly lower gap in performance between the sex groups. We attribute the difference in performance on the different fatty tissue types to the varying distributions in fat volume between the sex groups.

\begin{figure}
    \centering
    \includegraphics[width=0.9\textwidth]{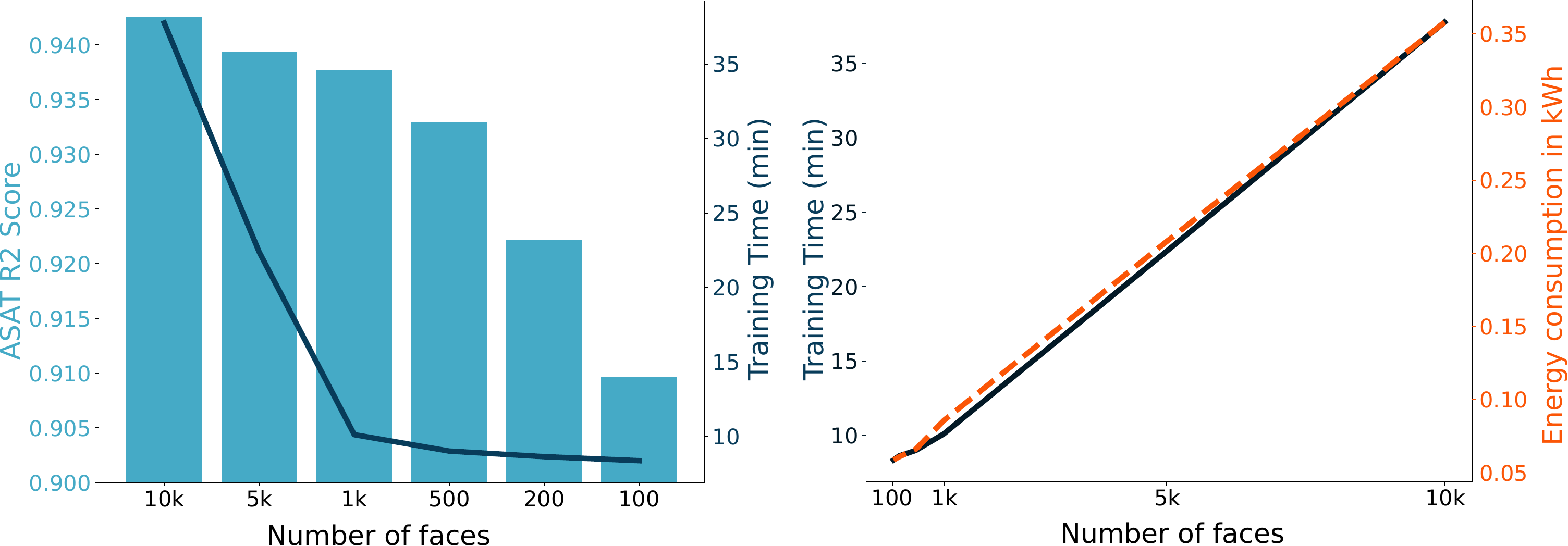}
    \caption{Relationship between training time and decimation rate of the meshes; The left plot shows the ASAT R2 scores (bars) and the corresponding training time, the right plot shows the linear relation between the training time or the energy consumption in kWh and the number of faces of the meshes.}
    \label{fig:runtimes}
\end{figure}

\section{Discussion and Conclusion}
In this work, we introduce a graph neural network-based method that enables adipose tissue volume prediction for visceral (VAT) and abdominal subcutaneous (ASAT) fat from triangulated surface meshes. The assessment of fatty tissue has high clinical relevance, since it has been shown to be a strong risk factor for diseases like type 2 diabetes and cardiovascular diseases \cite{kivimaki2017overweight,larsson2020body}. Especially a separate estimation of the two different fat tissues VAT and ASAT has shown to be a relevant medical assessment, since VAT is known to have a higher correlation with disease development compared to ASAT \cite{matsuzawa1995visceral,bergman2006visceral,shuster2012clinical}. 
We here use graph neural networks and triangulated surface meshes, extracted from full-body MR scans and show that they achieve accurate VAT and ASAT volume predictions.
We investigate how different decimation rates impact model performance and training times. Figure \ref{fig:runtimes} visualises this correlation. The bars in the left figure show the average ASAT volume prediction R2 scores on the test set of the GNNs trained on the differently decimated meshes. The overlaid line plot notes the corresponding training times. We can see that at one thousand faces, we reach an optimal trade-off between training time and performance. Training the GNN on the meshes with one thousand faces only takes about $10$ minutes and achieves high results of $0.893$ R2 on VAT and $0.935$ on ASAT volume prediction. On the right in Figure \ref{fig:runtimes}, we visualise the linear relation between the training time and the number of faces in the meshes. Training time also corresponds linearly to energy consumption in kWh. We attribute the comparably high performance of the strongly decimated meshes to the fact that the most outer coordinates/nodes still remain in the meshes, which carry a lot of information about the outline of a body.

The light-weight nature of GNNs allows for the usage of the full 3D data, while significantly reducing resource requirements and run time compared to 3D image-based methods. This shows great promise in the effort of bridging the gap between cheap, fast, but imprecise measures --such as BMI and waist circumference-- and time-consuming, costly, but accurate methods such as medical imaging (CT, MR, or DXA).

\section{Limitations and Future Work}
We see high potential in the utilisation of surface meshes and graph neural networks, given that the full 3D data can be utilised compared to only using binary silhouette projections like in \cite{klarqvist2022silhouette}. The low training times as well as the high scores of the GNNs show the successful application to fat volume prediction.
We note that we compare the run time of the training loops only. This does not include any pre-processing that is required for both silhouette-based and surface mesh-based approaches. 
The GNN architecture is based on SAGE graph convolutions \cite{hamilton2017inductive}, because they achieved the best results in our experiments, compared to graph attention networks \cite{velivckovic2017graph} and graph convolutional networks \cite{kipf2016semi}. A potential improvement of our method would be the utilisation of other mesh-specific convolutions such as adaptive graph convolution pooling \cite{gopinath2019adaptive} or FeaStNet \cite{verma2018feastnet}. 
Another interesting direction to explore is the utilisation of deeper GNNs. Li et al. \cite{li2019deepgcns}, for example, introduce a method that enables the utilisation of deeper GNNs without over-smoothing --a commonly known problem with GNNs. Over-smoothing refers to the issue that deep GNNs do not achieve high performance because all node embeddings in the graph converge to the same value \cite{li2018deeper}.

Our experiments are performed on surface meshes, that were extracted from MR images. However, we envision this method to work equally well on designated surface scans, without requiring expensive and time-consuming MR scans. We intend to investigate this in future work and apply our method to surface scans, which are for example acquired for dermatological examinations. This would eliminate the need for expensive MR scans and could lead to an embedding of this technique into routine medical examination.

\subsubsection{Acknowledgements} 
TM and SS were supported by the ERC (Deep4MI - 884622). This work has been conducted under the UK Biobank application 87802. SS has furthermore been supported by BMBF and the NextGenerationEU of the European Union.

\bibliographystyle{splncs04}
\bibliography{literature}

\end{document}